\documentclass[prl,twocolumn,showpacs,superscriptaddress]{revtex4-1}
\usepackage{graphicx,amssymb,amsmath,psfrag,color}
\begin{document}
\definecolor{darkgreen}{rgb}{0,0.5,0}
\newcommand{\be}{\begin{equation}}
\newcommand{\ee}{\end{equation}}
\newcommand{\jav}[1]{#1}

\title{Escort distribution function of work done and diagonal entropies in quenched Luttinger liquids}

\author{Bal\'azs D\'ora}
\email{dora@eik.bme.hu}
\affiliation{Department of Physics and BME-MTA Exotic  Quantum  Phases Research Group, Budapest University of Technology and
  Economics, 1521 Budapest, Hungary}

\date{\today}

\begin{abstract}
We study the escort probability distribution function of work done during an interaction  quantum quench of Luttinger liquids.
It  crosses over from
the thermodynamic to the small system limit with increasing $a$,  the order of the escort distribution, and 
depends on the universal combination $(|K_i-K_f|/(K_i+K_F))^a$ with $K_i$, $K_f$ the initial and final Luttinger liquid parameters.
From its characteristic function, the diagonal R\'enyi 
entropies and the many body inverse participation ratio (IPR) are determined to evaluate the information content of the time evolved wavefunction in terms
of the eigenstates of the final Hamiltonian. The  hierarchy of overlaps is dominated by that of the ground states.
The IPR exhibits a crossover from  Gaussian to power law decay with increasing interaction quench parameter.
\end{abstract}


\maketitle

Non-equilibrium dynamics plays an important role in many areas of contemporary physics, ranging from cosmology through condensed matter to cold atoms.
Beautiful theories  have been proposed and tested experimentally\cite{polkovnikovrmp,dziarmagareview}, focusing mostly on few-body observables.
However, deeper insights into a quantum system may be
gained by obtaining the full statistics of a given quantity.
In particular, the full distribution function of the interference contrast of coherently split one-dimensional Bose gas   was considered\cite{hofferberth,gring}, and
the theory of statistics of quantum work done during a time dependent process has been worked out\cite{rmptalkner,silva}.
However, does the full distribution function of  given observable contain all relevant information?

Given an original probability distribution $p_i$, its statistical and probabilistic attributes may be
scanned and revealed by studying the associated escort distribution\cite{beck}, defined as $P_i=p_i^a/(\sum_np_n^a)$, where  $a>0$ is the order of the escort distribution.
For $a > 1$ the escort distribution emphasizes the more likely events and suppresses the more
improbable ones.  For $0 < a < 1$, the escort distribution accentuates
less probable, rare events.
The introduction of escort distributions turns out to be useful in many areas of science (see \cite{beck,tsallis} and references therein) e.g. in nonextensive statistical 
mechanics, for analyzing earthquakes and structural degradation of matter, 
quantifying the efficiency of source coding in  information theory and the entropy in black holes, for the statistical analysis of financial data, describing fractals\cite{halsey} etc.

Escort distributions also facilitate the comparison of various probability distributions (PDs). In case of slow decay at infinity (e.g. Cauchy distribution),
the moments above a given one can diverge, and the usual characterization fails. However, escort distribution converge faster and
can provide well-defined quantities for the moments, which is the typical scenario
 within nonextensive statistical mechanics\cite{tsallis}.

The escort parameter is also understood as having $a$ replicas of a system and considering only those instances when all replicas are exactly 
in the same state $i$, which occurs
 with probability $p_i^a$. In some cases\cite{calabrese2009,stephan}, it is even more convenient to consider $a$ replicas of a system and 
calculate the $a$th power of probabilities.

Escort distribution can reveal additional information about quantum systems as well.
For example, the energy levels of electrons in a magnetic field form fractal structure, known as the Hofstadter's butterfly\cite{hofstadter}.
For a non-integrable quantum system, the level statistics deviate from Poisson distribution and become more Wigner-Dysonian\cite{emary}, indicating level repulsion. 
Such systems are expected to reveal quantum chaotic behaviour\cite{beck}, and might
possess complicated  PDs, whose hidden structures can be revealed by the escort PDs.

Recently much attention has been focused on the PD function of work done during a quantum quench and on the closely related Loschmidt 
echo\cite{deffner,silva,heylkehrein,dorner,mazzola,doraLE}.
Therefore, we investigate the  escort PD function of work done in a notoriously strongly correlated system, a Luttinger liquid (LL) after an interaction quench\cite{cazalillaprl}
and show that it is connected to the diagonal R\'enyi entropies\cite{polkovnikovannphys,santos}, where the diagonal elements of the 
density matrix in the instantaneous basis are used. 
A LL is realized in many one-dimensional fermionic, bosonic and spin systems\cite{giamarchi,nersesyan}. 
Although the Luttinger model is far from being non-integrable, it is useful to reveal the merit of 
focusing on the escort PD in this exactly solvable and physically relevant model, before departures from integrability are taken into account.

The escort PD of work done is
\begin{gather}
P_a(W)=\frac{1}{\sum_np_n^a} \sum_m p_m^a\delta(W-E_m), 
\label{escortpa}
\end{gather}
with $a> 0$, and 
\begin{gather}
p_m=|\langle m|G_0\rangle|^2.
\end{gather}
Here, $|G_0\rangle$ is the initial many body ground state wavefunction, while $|m\rangle$'s are the many body eigenstates of the final Hamiltonian, obtained after a quantum quench.

The corresponding escort characteristic function of the  unnormalized escort distribution is defined as
\begin{equation}
G_a(t)=\sum_m p_m^a \exp(iE_mt),
\label{gat}
\end{equation}
and  the PD from Eq. \eqref{gat} becomes normalized when $G_a(t)/G_a(0)$ is Fourier transformed, and by definition, $G_1(0)=1$.

A LL is  described by bosonic  sound-like collective excitations,
regardless to the  statistics of the original system. The LL
Hamiltonian is given by\cite{giamarchi}
\begin{equation}
H(t)=\sum_{q\neq 0}\left(  \omega_q^0 a_q^+ a_q
+\frac{g(q)\Theta(t)}{2}[a_qa_{-q}+a_q^+a_{-q}^+]\right) \;,
\label{eq:LL}
\end{equation}
where $g(q)=g_2 |q|$ with $g_2$ the strength of the quenched interaction, and $\omega^0_q\sim |q|$ the initial bosonic spectrum. 
Assuming $K_i$ and $K_f$ initial and final LL parameters\cite{giamarchi}, respectively, the relative LL parameter is $K=K_f/K_i$\cite{doraLE},
which determines the angle $\theta$ of Bogoliubov rotation  from the initial to the final Hamiltonian in equilibrium as
$\sinh^2(\theta)=(1-K)^2/4{K}$. The final state dispersion is
$\omega_q=v|q|$ with $v$  the sound velocity.

The wavefunction of a Luttinger liquid is known \cite{pham,fjaerestad,doraLE}.
The excited states are constructed by populating the bosonic vacuum.
Working in the basis of the final Hamiltonian (after a Bogoliubov rotation of $H(t>0)$ to render it diagonal), 
the eigenfunctions having a finite overlap with the initial state, are of the form
\begin{gather}
|m\rangle=\prod_{q>0}|n_q\rangle=\prod_{q>0} \frac{1}{n_q!}\left(b^+_qb^+_{-q}\right)^{n_q}|0\rangle
\end{gather}
i.e. having the same number of bosons in a given $q$ and $-q$ state. In this basis,
\begin{gather}
|G_0\rangle=\prod_{q>0}|G_0^q\rangle=
\prod_{q>0}\frac{1}{\cosh(\theta)}\exp\left(-\tanh(\theta)b^+_qb^+_{-q}\right)|0\rangle.
\end{gather}

From the specific structure of the bosonic wavefunction\cite{pham,fjaerestad,doraLE},
we get
\begin{gather}
G_a(t)=\sum_m |\langle m|G_0\rangle|^{2a}\exp(iE_mt)=\nonumber\\
=\prod_{q}\sum_{n_{q}}|\langle n_{q}|G_0^q\rangle|^{2a}\exp(iE_{n_{q}}t),
\end{gather}
and $\langle n_{q}|G_0^q\rangle=\tanh^{n_q}(\theta)/\cosh(\theta)$.
For a given mode,
the overlap is calculated as
\begin{gather}
\sum_{n}|\langle n_q|G_0^q\rangle|^{2a}\exp(iE_{n_q}t)=\frac{\cosh^{-2a}(\theta)}{1-\tanh^{2a}(\theta)\exp(2i\omega_q t)}
\end{gather}
where $E_{n_q}=2n_q\omega_q$,  and the factor of 2 comes from the pair of entangled boson modes at a given $\pm q$.
The numerator comes from the normalization factor, while the denominator accounts for the overlap of multiboson wavefunctions.
The escort characteristic function yields
\begin{gather}
G_a(t)=\prod_{q>0}\left(\cosh^{2a}(\theta)-\sinh^{2a}(\theta)\exp(2i\omega_q t)\right)^{-1}.
\end{gather}
After some algebra, it is evaluated in closed form using an exponential cutoff, $\exp(-\alpha|q|)$ for the bosonic modes as
\begin{gather}
\ln\left( \frac{G_a(t)}{G_a^\infty}\right)=\frac{L}{2\pi}\frac{\tanh^{2a}(\theta)}{\alpha-2ivt}\times\nonumber\\
\times  _3F_2\left(1,1,1+\frac{i\alpha}{2tv};2,2+\frac{i\alpha}{2tv};\tanh^{2a}(\theta)\right),
\label{G_a(t)}
\end{gather}
where $_3F_2(a;b;z)$ is the generalized hypergeometric function\cite{gradstein} and
\begin{gather}
G_a^\infty\equiv G_a(t\rightarrow\infty)=\left[\cosh(\theta)\right]^{-La/\pi\alpha},
\label{gainfty}
\end{gather}
which is the $2a$th power of the respective ground state overlaps, namely $[\cosh(\theta)]^{-L/2\pi\alpha}$, extending the result for $a=1$\cite{doraLE}.
This is the generalization of the many body orthogonality catastophe to the escort distribution case.

\begin{figure}[t!]
\psfrag{x}[t][][1][0]{$W/\overline W_a$}
\psfrag{y}[b][t][1][0]{$P_a(W)\overline W_a$}
\psfrag{xx}[t][][0.8][0]{$W/\overline W_a$}
\psfrag{yy}[b][t][0.8][0]{$P_a(W)\overline W_a$}
\psfrag{L=10}[][][1][0]{$L/2\pi\alpha=10$}
\psfrag{a=0.5}[][][0.9][0]{$\tanh^{2a}(\theta)=\frac 12$}
\includegraphics[width=7cm]{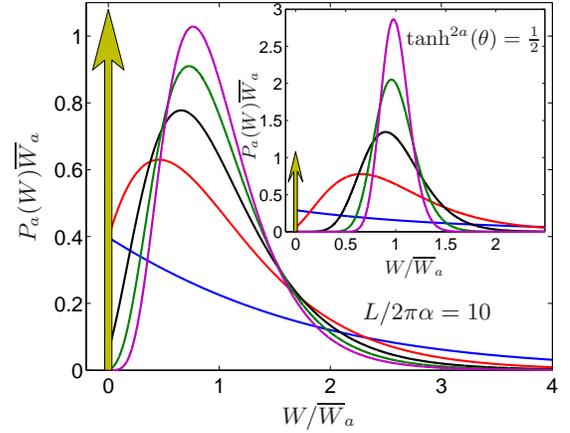}
\caption{The escort PD function of work is plotted for $L/2\pi\alpha=10$ and $\tanh^{2a}(\theta)=0.1$ (blue), 0.3 (red), 0.5 (black), 0.7 (green)
and 0.9 (magenta) with peak position from left to right. The vertical arrow at $W=0$ denotes the Dirac delta function with spectral weight $P_a^{ad}$, given in Eq. \eqref{pad}.
The inset shows the evolution of the $\tanh^{2a}(\theta)=0.5$ case for $L/2\pi\alpha=1$ (blue), 10 (red), 40 (black) 100 (green) and 200 (magenta). For the last four, the spectral
weight of the Dirac delta peak is practically zero.
\label{escortpdf}}
\end{figure}

The normalized escort PD is obtained from the characteristic function using Eq. \eqref{G_a(t)} as
\begin{gather}
\mathcal G_a(t)\equiv \frac{G_a(t)}{G_a(0)}=\left(1-\tanh^{2a}(\theta)\right)^{L/2\pi\alpha}\frac{G_a(t)}{G_a^\infty}.
\end{gather}
Based on this, we observe that both $\mathcal G_a(t)$  and the escort distribution of work $P_a(W)=\int\frac{dt}{2\pi}\exp(-iWt)\mathcal G_a(t)$ 
depend on the interaction and escort parameter through the combinations $\tanh^{2a}(\theta)$. 
Therefore, by varying the interaction strength and the escort parameter appropriately,
the escort work distribution function remains unchanged.
In addition,  the escort PD remains invariant under the $(K_i,K_f)\longleftrightarrow(K_f,K_i)\longleftrightarrow (1/K_i,1/K_f)\longleftrightarrow (1/K_f,1/K_i)$ changes of the 
initial and final Hamiltonian.

For $a=1$, we immediately get the characteristic function of work\cite{rmptalkner}, whose absolute value is the Loschmidt echo\cite{doraLE}.
The Fourier transform of $\mathcal G_1(t)$ gives the PD to find the system in a given energy eigenstate after the quench. However, it does not reveal
how many different eigenstates live on the same energy shell. The Fourier transform of $\mathcal G_{a\neq 1}(t)$
contains information about the number of states within a given energy shell as well, i.e. about degeneracies.
The escort distribution function of work done during the quench is visualized in Fig. \ref{escortpdf}.

These results are non-perturbative in the interaction strength, and agree   qualitatively with the perturbative,  non-escorted distribution of work done\cite{dorapdf}.
The finite probability to stay in the adiabatic ground state is
\begin{gather}
P_a^{ad}=\left[1-\tanh^{2a}(\theta)\right]^{L/2\pi\alpha},
\label{pad}
\end{gather}
signaled by the Dirac delta peak at zero energy and $P_a(W<0)=0$.
The $a$-escorted expectation value and variance of work follow from expanding $\ln \mathcal G_a(t)$ in $t$ as
\begin{subequations}
\begin{gather}
\overline W_a=\frac{Lv}{\pi\alpha^2}\frac{\tanh^{2a}(\theta)}{1-\tanh^{2a}(\theta)}\label{mean},\\
\sigma_W^2=\frac{4Lv^2}{\pi\alpha^3}\frac{\tanh^{2a}(\theta)}{(1-\tanh^{2a}(\theta))^2}\label{variance}.
\end{gather}
\label{meanvariance}
\end{subequations}
In the so-called small system limit\cite{rams}, defined by $L \tanh^{2a}(\theta)/2\pi\alpha\ll 1$, an exponential distribution with 
rate parameter $2v/\alpha$ accounts for the escort distribution, though most of the spectral weight is concentrated to the $W=0$ Dirac delta peak.
This is also corroborated by $\sigma_W/\overline W_a\xrightarrow{\tanh^{2a}(\theta)\rightarrow 0} \infty$.
In the opposite, 
thermodynamic limit ($L \tanh^{2a}(\theta)/2\pi\alpha\gg 1$), achievable by increasing $L$ or $\theta$ or decreasing $a$,
the distribution develops a sharp and narrow peak, centered at $\overline W_a$ and carrying almost all the spectral weight, 
as expected from the central limit theorem, since $\sigma_W/\overline W_a\xrightarrow{L\rightarrow\infty} 0$ from Eqs. \eqref{meanvariance}.
Around $\overline W_a$, there is a large number of degenerate overlaps with small individual probabilities.
In the extreme $a=0$ limit, all probabilities in Eq. \eqref{escortpa} become identical, and $P_0(W)$
yields the many-body density of states of a LL.

With increasing $a$, the large probability states are favoured, and the escort 
distribution approaches that in the small quench limit and the peak moves towards zero energy and disappears, and decays monotonically for larger energies.
This indicates that low total energy states are more similar to the initial states and appear with larger probabilities in the time evolved wavefunction.

In the opposite, decreasing  $a$ region, the escort distribution enhances the role of low probability  states and the number of states around a given energy determine the distribution.
Therefore, states with large total energy and large degeneracy overwhelm the smaller number of low energy states and dominate the distribution.

\begin{figure}[h!]
\psfrag{x}[t][][1][0]{$K_f/K_i$}
\psfrag{y}[b][t][1][0]{$S_a2\pi\alpha/L$}
\psfrag{ a=1/4}[][][0.8][0]{ $a=1/4$}
\psfrag{ a=1/2}[][][0.8][0]{ $a=1/2$}
\psfrag{ a=1}[][][0.8][0]{ $a=1$}
\psfrag{ a=2}[][][0.8][0]{ $a=2$}
\psfrag{ a=4}[][][0.8][0]{ $a=4$}
\psfrag{ a=inf}[][][0.8][0]{ $a=\infty$}
\includegraphics[width=7cm]{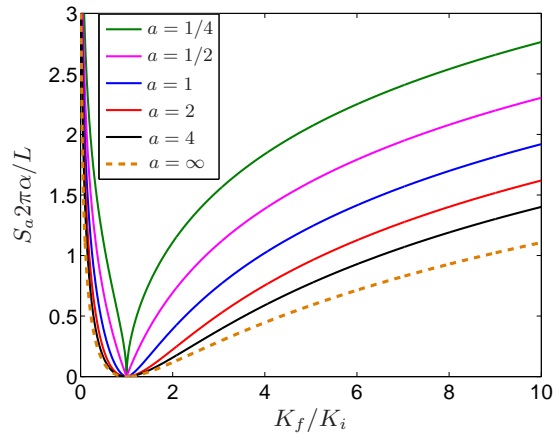}
\caption{The R\'enyi entropies as a function of $K=K_f/K_i$ for $a=1/4$, 1/2, 1, 2, 4 and $\infty$ (from top to bottom) are plotted.}
\label{entropies}
\end{figure}

The global information content of the quenched wavefunction in terms of the eigenstates of the final Hamiltonian is conveniently characterized by the 
diagonal von Neumann or  Shannon\cite{polkovnikovannphys,santos,collura} and  R\'enyi entropies what we obtain from $G_a(t)$ as well. 
Note that these characterize the information content of the original PD and not the escort ones, though the entropies of the escort PDs can also be
evaluated similarly.
Setting $t=0$, and using the definitions of the R\'enyi entropies, we get
\begin{gather}
S_a=\frac{1}{1-a}\ln\left(\sum_mp_m^a\right)=\frac{1}{1-a}\ln\left(G_a(0)\right)=\nonumber\\
=\frac{L}{2\pi\alpha}\frac{1}{a-1}\ln\left[\cosh^{2a}(\theta)-\sinh^{2a}(\theta)\right].
\end{gather}
The entropy is small when the probabilities are dominated by a few states, and grows with the number of final states contributing to the initial wavefunction.
Thus, it quantifies  entanglement and quantum fluctuations.
 
The largest probability, $p_{max}=\max_m p_m$ is the weight of the most probable configuration and is connected to the entropies as
\begin{gather}
p_{max}=\exp(-S_\infty)=\left(\frac{K_i+K_f}{2\sqrt{K_i K_f}}\right)^{-L/\pi\alpha},
\end{gather}
and it is identified as the probability to stay in the ground state, $|\langle 0|G_0\rangle|^2$ from $P_1^{ad}$ in Eq. \eqref{pad}, which dominates over the large number of low probability excited state overlaps.

While a direct computation of the von Neumann entropy would be rather difficult for the present case, similarly to other instances\cite{calabrese2009}, it follows from 
the R\'enyi entropies as the $a\rightarrow1$ limit as
\begin{gather}
S_1=\frac{L}{2\pi\alpha}\left(\cosh^2(\theta)\ln\cosh^2(\theta)-\sinh^2(\theta)\ln\sinh^2(\theta)\right),
\end{gather}
which plays the role of the thermodynamic entropy after the quench\cite{polkovnikovannphys}.
Various entropies as a function of the LL parameter are plotted in Fig. \ref{entropies}.
In the small quench limit ($K\approx 1$), it becomes a non-analytic function of $K$ for $a<1$ as
$S_a\sim |K-1|^{\min(2,2a)}$.
Let us note that $G_a(0)$ yields  also the non-extensive Tsallis entropies, which could become extensive for certain values of 
$a\neq 1$ for certain models as well\cite{caruso} (e.g. for the transverse field Ising chain). 
For the present situation, however all $a\neq1$ Tsallis entropies are non-extensive.

Another useful characteristics of the difference between the initial state  and the eigenstates of the final Hamiltonian is the many-body inverse participation 
ratio (IPR)\cite{neuenhahn,santosipr}, 
measuring
 the inverse  number of many body eigenstates of the final Hamiltonian over which the initial state is distributed. 
For the quenched
LL, it reads as
\begin{gather}
\textmd{IPR}=\sum_m p_m^2=G_2(0)=\left(\frac{K_i}{2K_f}+\frac{K_f}{2K_i}\right)^{-L/2\pi\alpha},
\end{gather}
and its logarithm, $S_2$ is plotted in Fig. \ref{entropies}. For small quenches, $K_f=K_i+\delta K$ with $|\delta K|\ll 1$,
the $\textmd{IPR}\approx \exp(-L(\delta K)^2/K_i^24\pi\alpha)$ decays as a Gaussian with the quench parameter. For sizeable quenches $K_f\gtrless K_i$,
 however, it crosses over to a power law decay $\textmd{IPR}\sim (K_f/K_i)^{-\textmd{sign}(K_f-K_i)L/2\pi\alpha}$ with respect to the LL parameter. 
These are roughly consistent with recent numerics\cite{neuenhahn}.
From the non-escorted, $a=1$ distribution of work done, 
we also determine that for small quenches, the contribution of low energy states in the expansion of the time evolved wavefunction is dominant over high energy ones.
With increasing quench size, however, the central limit theorem holds and most of the spectral weight comes from the large number of degenerate states located around the average energy.

Our calculations can be extended for higher dimensional and/or gapped bosonic systems as well. 
For example, quenching a one dimensional gapless system to a gapped phase, 
 the IPR decays exponentially with the gap $\Delta$ as
\begin{gather}
\textmd{IPR}=\exp\left(-{cL\Delta}/{v}\right)
\end{gather}
with $c>0$, and in particular, $c=(\sqrt 2-1)/4$ when quenching to the semiclassical limit of the sine-Gordon model\cite{iuccisinegordon}.
It would also be interesting to explore the behaviour of the escort distribution of work done in other models, for e.g. the Rabi model\cite{braak} and for local quenches such
as the X-ray edge problem\cite{knap}.

To conclude and answer the question raised at the beginning of the paper, the PD allows for calculating arbitrary expectation values of a given quantity, but it cannot resolve the interplay of degeneracies and individual probabilities.
An escort PD, on the other hand, is capable of revealing this additional information.
In addition to  demonstrate this for a Luttinger liquid, the diagonal R\'enyi entropies and the inverse participation ratio are shown to  follow also from the escort characteristic function of work done.

\begin{acknowledgments}

This research has been  supported by the Hungarian Scientific  Research Funds Nos. K101244, K105149, K108676, 
by the ERC Grant Nr. ERC-259374-Sylo and by the Bolyai Program of the HAS.
\end{acknowledgments}

\bibliographystyle{apsrev}
\bibliography{wboson}

\end{document}